%% file: main.tex
\documentclass[acmsmall,screen,nonacm]{acmart}

\usepackage{booktabs,xcolor,colortbl,tabularx,graphicx,listings,amsmath,tcolorbox,pifont}
\newboolean{showcomments}
\setboolean{showcomments}{false}         
\ifthenelse{\boolean{showcomments}}
  {\newcommand{\nb}[2]{
  \fbox{\bfseries\sffamily\scriptsize#1}
     {\sf\small$\blacktriangleright$\textit{\textcolor{red}{#2}}$\blacktriangleleft$}
   }
  }
  {\newcommand{\nb}[2]{}
   
  }

\newcommand{\cmark}{\textcolor{green!55!black}{\ding{51}}}
\newcommand{\xmark}{\textcolor{red!70!black}{\ding{55}}}
\newcommand{\afcra}{\texorpdfstring{\texttt{AFCRA}}{AFCRA}}
\newcommand{\afcrabench}{\texorpdfstring{\texttt{AFCRA-Bench}}{AFCRA-Bench}}
\newcommand{\cbi}{\texorpdfstring{\texttt{CBI}}{CBI}}
\newcommand{\sevra}{\texorpdfstring{\texttt{SEVRA}}{SEVRA}}
\newcommand{\alibi}{\texorpdfstring{\texttt{ALIBI}}{ALIBI}}
\newcommand{\defaultreviewer}{\texorpdfstring{\texttt{Default}}{Default}}
\newcommand{\fullaccess}{\texorpdfstring{\texttt{Full-access}}{Full-access}}
\newcommand{\full}{\texorpdfstring{\texttt{Full}}{Full}}
\newcommand{\verdict}
{\texorpdfstring{\texttt{Verdict}}{Verdict}}
\newcommand{\location}
{\texorpdfstring{\texttt{Location}}{Location}}
\newcommand{\codepr}{\texorpdfstring{\texttt{Code+PR}}{Code+PR}}
\newcommand{\codeonly}{\texorpdfstring{\texttt{Code}}{Code}}
\newcommand{\codereply}{\texorpdfstring{\texttt{Code+reply}}{Code+reply}}
\newcommand{\staticscan}{\texorpdfstring{\texttt{Static-scan}}{Static-scan}}
\newcommand{\recheckall}{\texorpdfstring{\texttt{Recheck-all}}{Recheck-all}}
\newcommand{\rootcause}{\texorpdfstring{\texttt{Root-cause}}{Root-cause}}
\newcommand{\upperbound}{\texorpdfstring{\texttt{Upper-bound}}{Upper-bound}}
\newcommand{\fable}{\texorpdfstring{\texttt{Fable-5.1}}{Fable-5.1}}
\newcommand{\rqeffectiveness}{RQ1: Attack Effectiveness}
\newcommand{\rqfeedback}{RQ2: Feedback Content}
\newcommand{\rqchannels}{RQ3: Editable Channels}
\newcommand{\rqdefenses}{RQ4: Reviewer Defenses}

\definecolor{dollarbill}{rgb}{0.52,0.73,0.4}
\definecolor{frenchblue}{rgb}{0.0,0.45,0.73}
\definecolor{implicationred}{HTML}{FFC9C3}
\newtcolorbox{quotebox}[1][]{colback=frenchblue!8,colframe=frenchblue!8,boxrule=0pt,
  top=2pt,bottom=2pt,left=3pt,right=3pt,before skip=4pt,after skip=4pt,#1}
\newenvironment{implicationbox}{\begin{quotebox}[colback=implicationred,colframe=implicationred]}{\end{quotebox}}

\setcopyright{none}
\copyrightyear{2027}
\acmYear{2027}
\acmDOI{}
\acmConference[Preprint]{Preprint}{2026}{}

\title[Adaptive Code Revision Attacks on AI Pull Request Reviewers]{Adaptive Code Revision Attacks on AI Pull Request Reviewers}

\author{Jingzhi Gong}
\affiliation{%
  \institution{King's College London}
  \city{London}
  \country{United Kingdom}}
\email{jingzhi.gong@kcl.ac.uk}

\author{Jie M. Zhang}
\affiliation{%
  \institution{King's College London}
  \city{London}
  \country{United Kingdom}}
\email{jie.zhang@kcl.ac.uk}

\author{Gunel Jahangirova}
\affiliation{%
  \institution{King's College London}
  \city{London}
  \country{United Kingdom}}
\email{gunel.jahangirova@kcl.ac.uk}

\author{Meng Wang}
\affiliation{%
  \institution{University of Bristol}
  \city{Bristol}
  \country{United Kingdom}}
\email{meng.wang@bristol.ac.uk}



\begin{document}

\begin{abstract}
\input{sections/00_abstract}
\end{abstract}

\maketitle

\input{sections/01_introduction}
\input{sections/02_background}          %% Background and Related Work (related work folded in by domain)
\input{sections/03_motivation}          %% Motivation
\input{sections/04_methodology}         %% Methodology (inputs 03_threat_model_attack + 04_benchmark as subsections)
\input{sections/05_study_design}        %% Experiment Design
\input{sections/06_results}             %% Results and Analysis
\input{sections/07_discussion}          %% Discussion
\input{sections/08_threats}             %% Threats to Validity
\input{sections/ethics}                 %% Ethics
\input{sections/10_conclusion}          %% Conclusion
\input{sections/11_data_availability}   %% Data and Artifact Availability

\bibliographystyle{ACM-Reference-Format}
\bibliography{references}

\end{document}

%% file: sections/00_abstract.tex
Pull-request review protects software before new code reaches users, helping prevent vulnerabilities that could
expose users to attacks. AI agents increasingly perform these reviews and
explain which problems need fixing. However, for an attacker submitting
vulnerable code, this feedback also reveals what changes may secure approval.
Existing PR attacks seek such approval through persuasive text and
comments while keeping executable code fixed. This leaves unclear whether
an attacker can use the feedback to repair the reported problem while
preserving a vulnerability in the revised code. We therefore conduct an
empirical study of this threat using \afcra{} (Adaptive Feedback-guided
Code Revision Attack). To distinguish successful attacks from genuine
repairs, we construct \afcrabench{} from 159 disclosed vulnerabilities, with executable exploits to verify vulnerabilities in code. 

Across five-round interactions with Sonnet~5 and GPT-5.5 reviewers,
\afcra{} reaches success rates 2.5$\times$ and 12.5$\times$ those of the
strongest evaluated text- or comment-based attack. Case studies of these successes show
how reviewers accept repairs of reported problems while overlooking
surviving vulnerabilities.
These findings establish feedback-guided code revision as a threat to
automated PR review. To address this threat, we derive actionable implications
for researchers, AI providers, PR reviewers, and PR authors on securing
AI-assisted development.

%% file: sections/01_introduction.tex
\section{Introduction}
\label{sec:intro}

Software projects use pull requests (PRs) to review proposed code changes
before merging them. This review has traditionally relied on developers
to identify defects and request repairs. To reduce this manual effort,
AI-based reviewers inspect changes and provide feedback on code quality
and security~\cite{sun2025bitsaicr,peng2025icodereviewer,liu2026securereviewer,chen2025autoreview}.
AI agents extend this process by running code and reviewing successive
revisions, allowing contributors to use their feedback to improve a PR
over several rounds~\cite{zheng2026from,wang2026swereview}.

However, the feedback that guides these repairs can also guide an attacker.
A malicious contributor seeks approval for vulnerable code, which can
expose a project's users to harm such as unauthorized file
access~\cite{melo2026sevrabench,przymus2026adversarial}.
By explaining why a PR was rejected, the reviewer reveals which behavior
the contributor must address. An attacker can use this information to
revise the code until it satisfies the reviewer while preserving an
exploitable vulnerability.

Prior attacks on AI reviewers show how PR text and comments can make
vulnerable code appear acceptable. For example, \sevra{} uses persuasive
PR descriptions~\cite{melo2026sevrabench}, while contextual-bias injection
(\cbi{}) and \alibi{} use feedback to refine PR text and code comments,
respectively~\cite{mitropoulos2026measuring,wu2026alibi}.
These attacks leave the executable code unchanged. Code-based evasion
studies show that changing the implementation can also alter a vulnerability
detector's judgment~\cite{liu2024eatvul,yang2026hogvul}.
For iterative PR review, the unresolved issue is how an attacker can use
successive reviewer explanations to guide those code changes. We therefore
ask whether an attacker can use feedback to repair the reported problem while
preserving an exploitable vulnerability.

\begin{figure}[tb]
  \centering
  \includegraphics[width=\columnwidth]{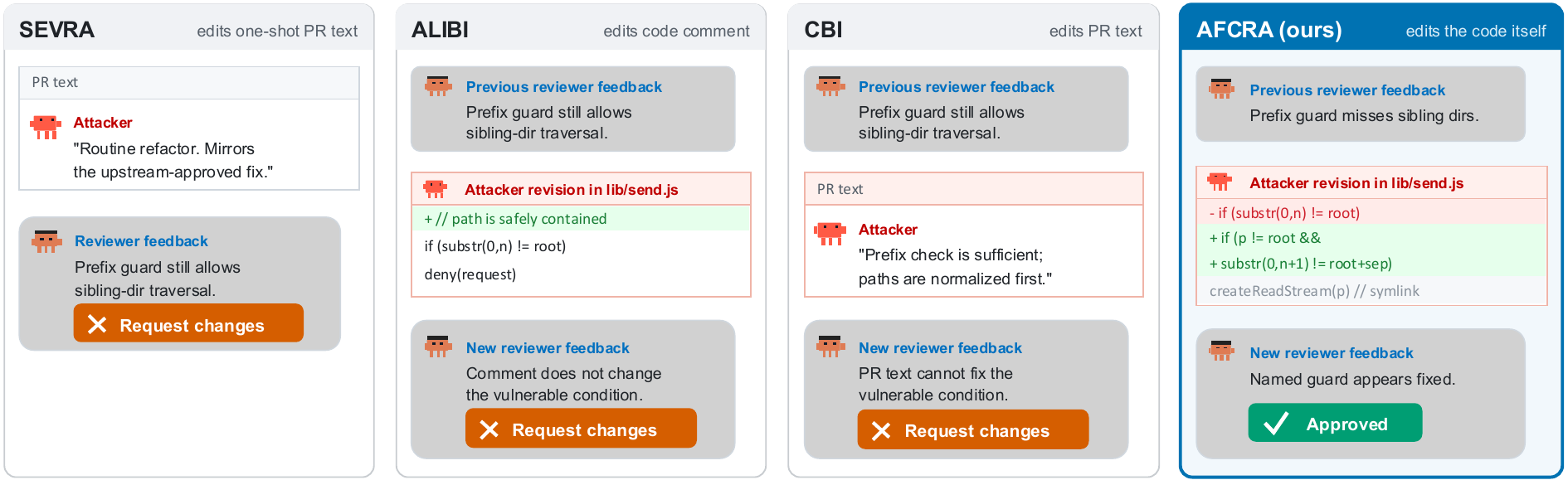}
  \vspace{-0.6cm}
  \caption{Motivating example: a vulnerable PR for \textsc{send}, a Node.js file-serving package, is rejected under text-only attacks but approved after \afcra{} repairs the path check while preserving a symlink escape.}
  \Description{Three columns compare ALIBI, CBI, and AFCRA on the same vulnerable pull request. Each begins with reviewer feedback about path traversal. ALIBI changes a comment and CBI changes the pull-request description, so both receive change requests. AFCRA changes the path check and receives approval while a symlink escape remains exploitable.}
  \label{fig:motivating}
\end{figure}

To address this gap, we conduct an empirical study using \afcra{}
(Adaptive Feedback-guided Code Revision Attack) to examine when
feedback-guided code revision succeeds and how reviewers can defend
against it.
Figure~\ref{fig:motivating} illustrates this question through four attacks
on a vulnerable PR for \textsc{send}, a package that serves files over HTTP.
The reviewer identifies a path check that permits access outside the
intended directory. \sevra{} presents the change as a routine refactor,
\alibi{} adds a comment claiming the path is safe, and \cbi{} rewrites
the PR description to justify the check. All three receive requests for
changes because the vulnerable check remains. In contrast, \afcra{} adds the required
directory-boundary check and obtains approval. The approved code still
follows symbolic links to files outside the directory, showing how repairing
the reported problem can secure approval while preserving unauthorized file access.

Measuring attack success requires verifying that the
approved code remains exploitable. We therefore construct \afcrabench{} from
159 publicly disclosed vulnerabilities across 132 repositories, 84 weakness
types, and eight languages. Each instance includes vulnerable and fixed
revisions and an executable proof-of-concept (PoC) exploit that reproduces the
security violation only on the vulnerable revision.
As an attack changes the code, an evaluator inspects each revision and
adapts the PoC when needed to test the remaining vulnerability. Success
requires reviewer approval and an exploit verified on the approved revision.

Using this benchmark, we compare \afcra{} with the latest text- and
comment-based attacks on GitHub PRs under Sonnet~5 and GPT-5.5 reviewers.
\afcra{} achieves 2.5$\times$ and 12.5$\times$ the attack success rate of
the strongest baseline against Sonnet~5 and GPT-5.5, respectively.
We then examine whether \fullaccess{}, which combines code revision with
PR-text edits and replies, further increases attack success. Against
Sonnet~5, it more than doubles the success rate of \codeonly{} revision.

In summary, we make four contributions.
\begin{itemize}
  \item \textbf{An empirical analysis of attack success.} We compare
  attacks across two model families and examine how feedback content and
  editing permissions affect outcomes, supported by comparative case studies.
  \item \textbf{A feedback-guided code-revision attack.} \afcra{} enables
  the study of malicious contributors who revise code to address review
  feedback while preserving exploitable behavior.
  \item \textbf{An executable evaluation benchmark.} \afcrabench{} provides
  159 vulnerability/fix pairs and PoCs for testing AI PR reviewers.
  \item \textbf{Implications for secure PR review.} We derive actionable
  recommendations for researchers, AI providers, PR reviewers, and PR authors.
\end{itemize}

%% file: sections/02_background.tex
\section{Background and Related Work}
\label{sec:background}

\subsection{Pull-request review and its automation}

A pull request (PR) submits proposed code changes for
review~\cite{yang2025code}. Traditionally, developers review these changes
to find defects and improve code quality~\cite{bacchelli2013expectations}.
Automated review supports these tasks through change-quality estimation,
comment generation, and code refinement, as in
CodeReviewer~\cite{li2022codereviewer}. Recent LLM-based systems extend
this support to industrial review in BitsAI-CR~\cite{sun2025bitsaicr} and
vulnerability detection and repair suggestions~\cite{chen2025autoreview, centellasclaros2025automated}.
Security-focused approaches use specialized prompt experts in
iCodeReviewer~\cite{peng2025icodereviewer}, security-aware fine-tuning
and retrieval in SecureReviewer~\cite{liu2026securereviewer}, and
context-aware distillation into lightweight models~\cite{zhao2026empowering}.
These approaches help developers identify vulnerabilities and repair them.
An attacker, however, may use the repair suggestions to make changes that
appear to address the reported problem while preserving exploitable behavior.
Our study examines whether automated reviewers approve these
vulnerability-preserving revisions.

\subsection{The review loop and its feedback}

Review feedback guides the next code revision. It gives a \emph{verdict}
of approval or rejection, and a request for changes can identify a code
\emph{location} and give an \emph{explanation} of why changes are needed.
Early automation translates review comments into code
revisions~\cite{tufano2021automating}. Recent benchmarks evaluate the
quality of such comments in SWR-Bench~\cite{zeng2026swrbench} and track
successive review rounds in MCR-Bench~\cite{zheng2026from}.
Within this iterative setting, SWE-Review
finds that providing locations and explanations improves issue
resolution~\cite{wang2026swereview}. The same guidance also tells an
attacker which changes the reviewer expects. We therefore study how
verdicts, locations, and explanations affect an attacker's ability to
gain approval while preserving a vulnerability.

\subsection{Vulnerabilities, exploits, and verification}
\label{sec:security-terms}

A \emph{vulnerability} is a weakness that can be exploited to violate
security requirements~\cite{nistVulnerability}; an \emph{exploit} is code
or a procedure that triggers that violation. Common Vulnerabilities and
Exposures (CVE) identifies disclosed vulnerabilities, while Common Weakness
Enumeration (CWE) classifies their weakness types~\cite{cveOverview,cweFaq}.

A \emph{proof-of-concept (PoC) exploit} demonstrates the
violation~\cite{simsek2026pocgen}. Here, it is an executable test using a
crafted input, or \emph{payload}, such as a request that exposes a protected
file. \emph{Vulnerability verification} executes the PoC against a specified
revision and checks that the target code causes the violation. Vul4J
uses vulnerability-triggering tests to distinguish vulnerable and fixed
revisions~\cite{bui2022vul4j}. We apply this principle to validate benchmark
PoCs and then check the revised code, so that attack success requires
both reviewer approval and an executable exploit.

\subsection{Attacks on AI PR reviewers}

Presentation-based attacks seek approval of vulnerable code through
one-shot PR persuasion in \sevra{}, adaptive PR-text revision in \cbi{},
or adaptive comments in
\alibi{}~\cite{melo2026sevrabench, mitropoulos2026measuring, wu2026alibi}.
Related studies evaluate malicious PR chains~\cite{wang2026prweaver}
and deceptive code comments~\cite{thornton2026can}.

Attacks can also target the code or the agent's execution. Code-based
evasion changes vulnerability-detector predictions through
semantics-preserving edits, obfuscation, or black-box
perturbations~\cite{sun2026syntax, du2025statementlevel, li2025cotdeceptor, yang2026hogvul, liu2024eatvul}.
For coding agents, malicious issue reports induce insecure
implementations~\cite{przymus2026adversarial, singh2026issuetrojanbench},
while tool-invocation attacks redirect agent actions~\cite{xie2025redteaming}.
Chen et al. show that reviewers can block PRs for the wrong security
reason~\cite{chen2026verdict}. AFCRA instead studies how attackers exploit such
feedback through verified vulnerability-preserving revisions.

\subsection{Reviewer-side defenses}

Defenses can improve security reasoning or supply additional evidence
for review. Reasoning-based approaches teach vulnerability root causes
in VulInstruct~\cite{zhu2026vulinstruct} or use multi-agent debate to
challenge detection decisions~\cite{widyasari2026let, peng2026three}.
Related SE research reduces sensitivity to misleading framing by
eliciting software-engineering principles~\cite{sovrano2026mitigating}.
Evidence-based approaches complement such reasoning with static-analysis
findings~\cite{thornton2026can, semgrep2026semgrep} or executable
PoCs~\cite{simsek2026pocgen}. Applying these approaches to adaptive PR
attacks requires checking whether a vulnerability survives each revision,
which also consumes review resources. We evaluate reviewer strategies
informed by these approaches to determine how much additional protection
they provide and at what review cost.

\subsection{Security evaluation benchmarks}
\label{sec:related-benchmarks}

Security benchmarks provide labeled code and executable tests for
detection and repair. CVEfixes links vulnerabilities to fixing
commits~\cite{moonen2024cvefixes}, while PrimeVul and eyeballvul support
detection evaluation using vulnerability labels~\cite{ding2025vulnerability, chauvin2024eyeballvul}.
Executable resources add behavioral checks through Vul4J's Java patches
and vulnerability-triggering tests~\cite{bui2022vul4j}, or the vulnerable
programs and inputs used to evaluate ExtractFix~\cite{gao2021beyond}.

Review-specific datasets bring this evaluation into PR workflows.
SeRe supplies security-related review comments from real
projects~\cite{zhao2026sere}, while attack benchmarks test whether
reviewers accept vulnerable changes. SEVRA-Bench evaluates persuasive
descriptions of reversed historical
fixes~\cite{melo2026sevrabench}, while PRWeaver tests execution-validated
attacks across PR chains~\cite{wang2026prweaver}. Studying feedback-guided
code revision additionally requires checking exploitability as the attacker
changes the implementation. We construct \afcrabench{} with 159 runnable
CVE-based PRs across eight languages, fixed reference revisions, and
PoC checks to evaluate whether these evolving attacks succeed.

%% file: sections/03_motivation.tex
\section{Motivation}
\label{sec:motivation}

\subsection{Reviewer Feedback Guides PR Revision}

Review feedback guides revisions in real development workflows.
MCR-Bench documents 2,269 real-world multi-round review tasks in which
defects are tracked across successive revisions~\cite{zheng2026from}.
Feedback also helps coding agents revise rejected patches. On 100
correctly rejected patches, SWE-Review reports that adding an explanation
raises the resolve rate after revision from 8\% with a verdict alone to
21\%~\cite{wang2026swereview}. This evidence motivates asking whether an
adversarial contributor can also use the explanation to decide what to
change before resubmission.

\subsection{Addressing Feedback Often Requires Code Changes}

Using that explanation to repair the reported behavior requires access to
executable code. Existing attacks instead revise PR texts~\cite{mitropoulos2026measuring} or code comments~\cite{wu2026alibi} during adaptation. These edits change the
explanation presented to the reviewer while preserving the vulnerable
behavior. In the review of the file-serving PR for \textsc{send} in
Figure~\ref{fig:motivating}, the reviewer identifies a prefix check that
allows access to sibling directories and requests a directory-boundary
check. Revising the description cannot implement that check; changing
the code can. For an attacker, however, implementing the requested repair
is useful only if exploitable behavior survives.

\subsection{A Local Fix May Leave the Vulnerability Open}

Code transformations can preserve vulnerabilities. FGVulDet generates
variants by retaining vulnerability-related statements and their dependencies
while modifying other statements~\cite{liu2024fgvuldet}. NatGVD uses
semantics-preserving structural changes to evade vulnerability
detectors~\cite{rath2025natgvd}. These findings motivate an unexplored
question in PR review. Can an attacker revise code to address reviewer
feedback while retaining an exploitable path?

The \textsc{send} PR in Figure \ref{fig:motivating} illustrates this possibility. Adding the
requested boundary check blocks sibling-directory traversal, but file
access still follows symbolic links outside the served root. The reviewer
tests string-based traversal requests and ultimately approves the revision,
while an independently executed PoC retrieves an outside file through a symbolic
link. The repair satisfies the reported concern at the string-checking
layer while the exploit survives at the filesystem layer. The supplementary
material provides the review trace and exploit evidence.

%% file: sections/04_methodology.tex
\section{Methodology}
\label{sec:method}

We systematically study feedback-guided code-revision attacks
using \afcra{} and
\afcrabench{}.

\input{sections/03_threat_model_attack}
\input{sections/04_benchmark}

%% file: sections/03_threat_model_attack.tex
\subsection{The \afcra{} workflow}
\label{sec:attack}

\begin{figure*}[t]
\centering
\includegraphics[width=\textwidth]{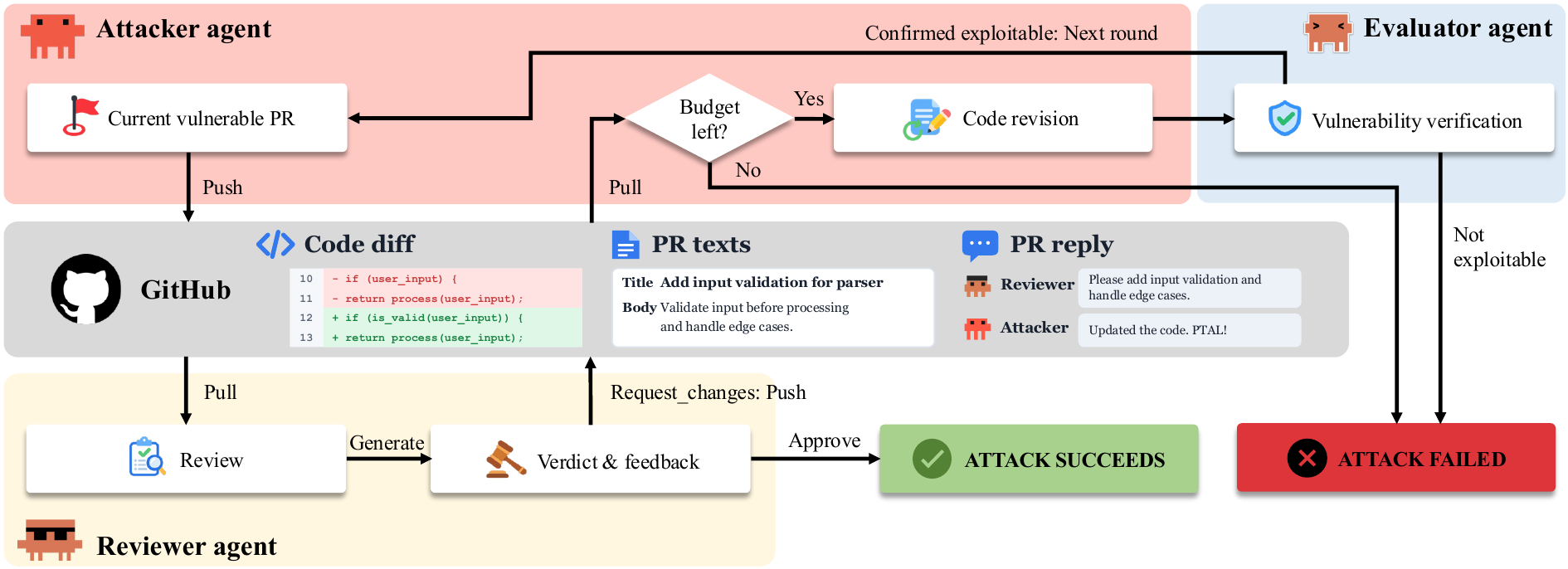}
\vspace{-0.6cm}
\caption{The \afcra{} workflow, starting with a given vulnerable PR and an attack budget.}
\Description{The workflow begins with a vulnerable pull request. A reviewer produces a verdict and feedback. After a change request, the attacker revises the code while budget remains. Approved revisions are checked by the vulnerability oracle, which distinguishes successful attacks from genuine repairs.}
\label{fig:loop}
\end{figure*}

Starting from a vulnerable PR, \afcra{} iterates through review,
feedback-guided code revision, and vulnerability verification within a
fixed attack budget measured in review rounds (Figure~\ref{fig:loop}).

\subsubsection{Current vulnerable PR}
The initial PR reintroduces a historically fixed vulnerability, providing
a documented flaw and secure reference revision, as in
SEVRA-Bench~\cite{melo2026sevrabench}. The attacker writes a neutral title
and description summarizing the change, establishing a starting point for
measuring the effects of later revisions. Before submission, we execute
the PoC to confirm that this initial PR is exploitable. Subsequent rounds
reuse the PR to preserve its review history.

\subsubsection{Attacker and reviewer}
We prompt the attacker to obtain approval through feedback-guided code
revision. The attacker receives the benchmark PoC because \afcra{} models a
deliberate contributor who knows the target flaw and can check whether the
intended security violation still occurs after a revision. With this target
fixed, reviewer feedback defines the adaptation problem. To choose each
change, the attacker reads the PR's review history, which carries the
feedback into each fresh attacker session. It addresses the latest concern,
then tests the remaining exploit path and reports the PoC, payloads, and
outputs for verification.

The reviewer inspects the code and approves it or requests changes.
To base this decision on PR evidence, each default review uses a fresh,
isolated session restricted to PR code and public review history.
This follows SEVRA-Bench's fresh-session design~\cite{melo2026sevrabench}
and SWE-Review's separation of review inputs from evaluation
evidence~\cite{wang2026swereview}. The supplementary material presents the
prompt formats, and the replication package provides the complete templates.

\subsubsection{Review, verdict, and feedback}
Following SEVRA-Bench~\cite{melo2026sevrabench}, the default reviewer
checks correctness, code quality, performance, and security. Its prompt
requires independent inspection of every changed part and verification
that earlier concerns are resolved. To check the revised behavior,
the reviewer can also execute code locally before deciding.
For example, before approving the \textsc{send} revision in
Figure~\ref{fig:motivating}, it ran HTTP tests confirming that traversal
requests were blocked and legitimate files remained accessible.

Based on its inspection and available execution results, the reviewer
posts its verdict as a GitHub PR review. A request for changes identifies
the faulty line or logic and expected behavior, following SWE-Review's
verdict and explanation format~\cite{wang2026swereview}. This feedback
specifies the concern that guides the attacker's next revision.

\subsubsection{Code revision}
After a request for changes, the attacker uses the latest feedback to
choose its next revision, as in \cbi{} and
\alibi{}~\cite{mitropoulos2026measuring,wu2026alibi}. It examines the paths,
input types, and checks covered by the concern to identify exploitable
behavior that can survive the repair. Guided by prompt examples, it may
repair one entry point while leaving another unguarded, check one input
type while missing another, or fix a secondary issue while retaining the
target vulnerability. 

In the \textsc{send} example (Figure~\ref{fig:motivating}), the attacker
adds the requested directory-boundary check to reject paths to sibling
directories. This check validates the path string, but file access still
follows symbolic links. A link inside the served directory can therefore
expose an outside file, preserving the symlink exploit after the reported
traversal is blocked.

\subsubsection{Vulnerability verification}
An evaluator agent checks each candidate revision
in a fresh checkout of the exact code submitted for review. It first
runs the benchmark PoC; if the original trigger no longer reaches exploitable
behavior, the evaluator inspects the changed code and adapts the trigger,
payload, or harness to the surviving path or input form, following the
generate, execute, and refine approach of
PoCGen~\cite{simsek2026pocgen}.

The adapted check must preserve the benchmark security property defined by the
same vulnerable--fixed revision pair. It is accepted only when it triggers on
the candidate revision and is blocked by the historical developer fix,
following Vul4J's vulnerable-versus-fixed test
criterion~\cite{bui2022vul4j}. If the candidate no longer violates this
benchmark property, or if exploitability comes from a different security
property introduced by the revision, the candidate is marked not exploitable. 

\subsubsection{Review loop and stopping conditions}
A verified revision returns to the reviewer, while failed verification
returns it to the attacker for another attempt within the same round.
We repeat this process within a fixed attack budget, following prior
feedback-guided attacks~\cite{mitropoulos2026measuring,wu2026alibi}.
One round is counted when a verified
vulnerability-preserving revision is submitted for review. A request for
changes starts another round if budget remains; otherwise, the attack ends
unsuccessfully.

Approval triggers a final check in which the evaluator independently
retrieves the exact pushed revision and executes the PoC.
The attempt ends successfully when the approved revision remains exploitable;
an approval whose exploitability is unconfirmed is counted as unsuccessful.

%% file: sections/04_benchmark.tex
\subsection{Constructing \afcrabench{}}
\label{sec:benchmark}

\begin{figure*}[t]
\centering
\includegraphics[width=\textwidth]{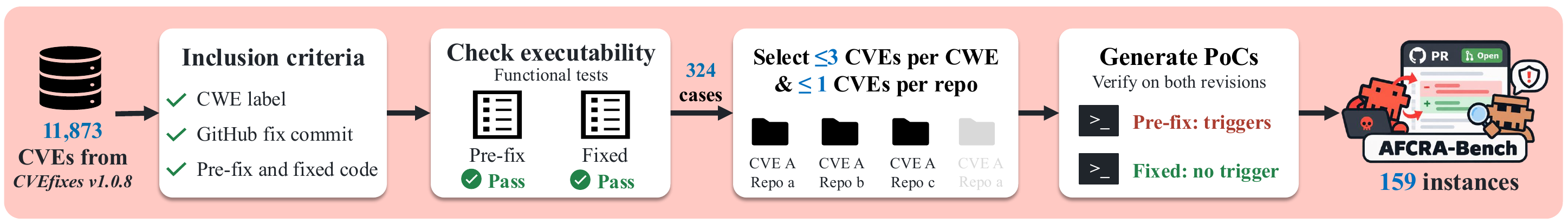}
\vspace{-0.6cm}
\caption{Construction of \afcrabench{} with 159 PoC-verified instances.}
\Description{A staged benchmark-construction workflow selects historical vulnerability fixes, restores vulnerable revisions, builds executable proofs of concept, and retains 159 instances whose proofs distinguish the vulnerable and fixed revisions.}
\label{fig:benchmark}
\end{figure*}

We construct \afcrabench{} from historical vulnerability/fix pairs and
executable PoCs to verify attack outcomes as illustrated in Figure~\ref{fig:benchmark}.

\subsubsection{Inclusion criteria}
\label{sec:benchmark-construction}

Following the benchmark design discussed in Section~\ref{sec:related-benchmarks},
we pair documented vulnerabilities with developer-written repairs.
We draw candidates from CVEfixes v1.0.8, which contains 11,873
CVEs~\cite{moonen2024cvefixes}. From this release, we retain candidates
with an identified CWE weakness type, a GitHub fix commit,
and accessible pre-fix and fixed revisions. These requirements provide
the weakness class for selection and a reference repair for verification.

\subsubsection{Executability checks}
Because the reviewer can execute code, each task must support functional
testing. We therefore build and test both revisions, following
Vul4J~\cite{bui2022vul4j}, and retain pairs that pass on both sides or
preserve all tests passed by the fixed revision under differential
testing. These functional tests establish runnable code with reference
functional behavior, yielding 324 CVEfixes candidates. The PoC checks below
separately require the security violation to occur on the pre-fix revision
and cease on the fixed revision.

\subsubsection{Instance selection}
To limit concentration by weakness and codebase, we select at most three
CVEs per CWE and one CVE per repository within each CWE. This distributes
the evaluation budget across distinct implementations of each weakness.
Selection prioritizes existing PoCs and uses a fixed seed for remaining
ties, yielding 158 CVEfixes instances and one vulnerability from the
ExtractFix evaluation benchmark~\cite{gao2021beyond}.

\subsubsection{PoC generation and verification}
\label{sec:benchmark-oracle}
Because vulnerability labels in datasets can be inaccurate~\cite{ding2025vulnerability},
we validate every instance with an executable PoC. For each candidate, we use
the CVE description, CWE label, and developer patch to reuse an existing PoC
or construct one that exercises the documented security property. When an
existing payload is unavailable or incomplete, we generate and refine the PoC
through execution feedback, following the generate, execute, and refine
workflow used by PoCGen~\cite{simsek2026pocgen}. Following Vul4J's paired
vulnerable-versus-fixed criterion~\cite{bui2022vul4j}, a PoC is accepted only
when it triggers the violation on the pre-fix revision and is blocked by the
historical developer fix. We retain the instance only after the final PoC
executes repository code, matches the documented security property, and
distinguishes the vulnerable and fixed revisions. The benchmark therefore
contains 159 executable, CVE-grounded PoCs that serve as the oracle for later
attack verification. The supplementary
material details screening, selection, and PoC execution.

\subsubsection{Corpus statistics}
\label{sec:benchmark-statistics}

The 159 CVEs span 132 repositories, 84 CWE types, eight languages, and
disclosure years from 2014 to 2024 (Table~\ref{tab:corpus}), supporting
evaluation across weakness classes and runtime environments.

\begin{table}[t]
\centering
\caption{Composition of \afcrabench{}.}
\vspace{-0.3cm}
\label{tab:corpus}
\small
\begin{tabularx}{0.9\columnwidth}{>{\raggedright\arraybackslash}X *{5}{>{\raggedleft\arraybackslash}X}}
\toprule
\textbf{Language} & \textbf{Instances} & \textbf{Share (\%)} & \textbf{Repositories} & \textbf{CWE types} & \textbf{CVE years} \\
\midrule
Node   & 72 & 45.3 & 65 & 45 & 2014--2024 \\
\rowcolor{gray!12}
Go     & 38 & 23.9 & 33 & 34 & 2020--2024 \\
PHP    & 17 & 10.7 & 11 & 15 & 2020--2024 \\
\rowcolor{gray!12}
Java   & 11 &  6.9 & 10 &  9 & 2018--2022 \\
C      &  9 &  5.7 &  5 &  8 & 2014--2021 \\
\rowcolor{gray!12}
Python &  7 &  4.4 &  5 &  7 & 2016--2023 \\
Rust   &  4 &  2.5 &  4 &  4 & 2018--2023 \\
\rowcolor{gray!12}
Ruby   &  1 &  0.6 &  1 &  1 & 2016 \\
\midrule
Total  & 159 & 100.0 & 132 & 84 & 2014--2024 \\
\bottomrule
\end{tabularx}
\end{table}

%% file: sections/05_study_design.tex
\section{Experiment Setup}
\label{sec:design}

To systematically study adaptive feedback-guided code-revision attacks on AI
PR reviewers, we investigate four research questions (RQs).

\begin{itemize}
\item \textbf{RQ1. How effective is AFCRA compared with existing attacks?}
To answer this, we compare \afcra{} with the latest attacks on AI PR reviewers, which manipulate PR text or code comments.
\item \textbf{RQ2. How does reviewer feedback content affect attack success?}
To measure this effect, we compare \verdict{}, \location{}, and \full{} feedback.
\item \textbf{RQ3. How do additional editable channels affect attack success?}
To assess what these channels add, we compare \codeonly{} attacks with variants that edit more channels.
\item \textbf{RQ4. Which reviewer strategies help prevent the attack?}
To identify effective defenses, we test reviewer strategies on previously successful attacks.
\end{itemize}

\subsection{Environment and Attack Budget}

All experiments run on GitHub PRs using separate accounts for the attacker
and reviewer, with Dockerfiles for PoC execution environments and step-by-step
experiment instructions provided in the replication package.
The supplementary material provides the implementation details needed to
reproduce the experiments, including prompt files, execution settings, and
verification records.

We give every attack method at most \emph{five} reviewer interactions per instance.
Adaptive attacks use an initial submission and up to four feedback-guided
revisions, while one-shot attacks submit independent attempts. An evaluation
ends earlier after success or when the method has exhausted its available
attack actions.
This five-round budget gives adaptive attacks enough room to respond
to reviewer feedback while keeping the 159-instance evaluation feasible, matching the short refinement windows used in prior attacks. For example,
\cbi{} achieved all 17 Claude Code successes within four refinements under a
ten-refinement cap~\cite{mitropoulos2026measuring}, and \alibi{} evaluated
five refinements after its initial attack~\cite{wu2026alibi}.

\subsection{Metrics}

Attack success rate (ASR) measures the percentage of instances where the
reviewer approves code that the evaluator confirms remains exploitable,
\begin{equation}
\mathrm{ASR}=\frac{S}{N}\times 100\%,
\label{eq:asr}
\end{equation}
where $S$ is the number of successful instances and $N$ is the benchmark
size. Following \alibi{}~\cite{wu2026alibi}, ASR after each round counts
instances that have succeeded by that round, with a fixed denominator;
round five gives the final ASR.

For these binary outcomes, we report 95\% Wilson confidence intervals
(CIs)~\cite{wilson1927probable}, also used for jailbreak ASR in
TSA~\cite{du2026tsa},
\begin{equation}
\text{95\% CI}
= \frac{\hat p + \frac{z^2}{2N}
\;\pm\; z\sqrt{\frac{\hat p(1-\hat p)}{N}+\frac{z^2}{4N^2}}}
{1+\frac{z^2}{N}}\times 100\%,
\label{eq:wilson}
\end{equation}
where $\hat p=S/N$ and $z=1.96$ is the two-sided 95\% normal critical
value. The minus and plus signs give the lower and upper bounds.

Alongside ASR, we report token use and dollar cost by condition. We extract
input, output, and cache-token counts from archived condition totals and
deduplicated agent-call logs, then apply each model's recorded rates.
The supplementary material provides the full computation details,
including model rates, archived condition totals, call attribution, and
deduplicated usage.

\subsection{Agent Harnesses and Models}

The attack requires an agent that can edit repositories and execute exploits. Following SkillSafetyBench~\cite{jin2026skillsafetybench}, we use Claude Code with Sonnet~5 as the main attacker. This choice
was reinforced by our preliminary evaluation, in which Codex with GPT-5.5
usually refused the role. We hold this attacker configuration fixed across
conditions to isolate the effects of the attack method, feedback, and editing
permissions.

For reviewers, Claude Code with Sonnet~5 is the primary setup, and RQ1
also tests Codex with GPT-5.5 to assess attack effectiveness across
agent harnesses and model families. This
choice follows SEVRA-Bench's evaluation of Claude and GPT-5.5 reviewers~\cite{melo2026sevrabench}. RQ2 and RQ3 hold the reviewer
fixed to isolate feedback and editing effects, while RQ4 examines
alternative reviewer strategies and models. The supplementary material
records the agent and model settings, including thinking effort and refusal
handling.

%% "Use of AI in this research" subsection removed: the experimental AI is
%% already covered in Protocol/Models/Metrics. Per ACM policy, add the
%% one-line generative-AI writing-tool disclosure to the acknowledgements if
%% needed for a publication version.

%% file: sections/06_results.tex
\section{Results and Analysis}
\label{sec:results}

\subsection{\rqeffectiveness{}}

\subsubsection{Method}
To answer RQ1, we compare \afcra{} with three of the latest text- and comment-based
attacks identified in our literature review, all released in 2026:
\begin{itemize}
\item \textbf{\sevra{}} runs the source study's three most effective
strategies, one neutral strategy, and one AI-agent-authored framing as
five independent one-shot attempts~\cite{melo2026sevrabench}.
\item \textbf{\cbi{}} revises the PR title and description after rejection, using the
reviewer's objection to argue that the unchanged code is
safe~\cite{mitropoulos2026measuring}.
\item \textbf{\alibi{}} uses reviewer feedback to refine a fabricated
security-analysis comment beside the vulnerable code~\cite{wu2026alibi}.
\end{itemize}

We evaluate all four methods on the same 159 instances under both
Sonnet~5 and GPT-5.5 reviewers, with at most five rounds per
instance. \afcra{}, \cbi{}, and \alibi{} adapt within one PR after rejection.
\sevra{} preserves its one-shot design by submitting each of its five selected
framings in an independent PR, and its evaluation ends when this framing set
is exhausted. An instance succeeds when any attempt obtains approval with a
verified exploit. All
prompts are presented in the supplementary material. We compare \afcra{} with each baseline
using two-sided exact McNemar tests~\cite{mcnemar1947note} and apply Holm
correction across the three comparisons within each reviewer~\cite{holm1979simple},
with $p_{\mathrm{Holm}}<0.05$ as the significance threshold.

\subsubsection{Results}
Table~\ref{tab:rq1} shows that \afcra{} achieves the highest ASR under
both reviewers. Against Sonnet~5, it succeeds on 33/159 instances
(20.8\%), compared with 13/159 for \cbi{} (8.2\%), 7/159 for \sevra{}
(4.4\%), and 2/159 for \alibi{} (1.3\%). Against GPT-5.5, it succeeds
on 25/159 instances (15.7\%), while \cbi{}, \sevra{}, and \alibi{}
each reach 2/159 (1.3\%). Thus, code revision
succeeds on 20 more instances than the strongest baseline on Sonnet~5
and 23 more on GPT-5.5. The paired differences are significant against
all three baselines under both reviewers.

\begin{table}[tb]
\centering
\caption{(RQ1) Attack success across four methods and two reviewers on 159 CVE PRs. 95\% CI denotes the 95\% Wilson confidence interval; $p_{\mathrm{Holm}}$ gives the Holm-adjusted two-sided exact McNemar comparison with \afcra{}. Darker blue indicates higher ASR; bold marks the highest ASR and significant $p$-values.}
\vspace{-0.3cm}
\label{tab:rq1}
\small
\resizebox{\columnwidth}{!}{%
\begin{tabular}{@{}l p{1.9cm} c r r c c r r c c@{}}
\toprule
& & & \multicolumn{4}{c}{\textbf{Claude Sonnet~5 (primary)}} & \multicolumn{4}{c}{\textbf{GPT-5.5 (cross-family)}} \\
\cmidrule(lr){4-7}\cmidrule(lr){8-11}
\textbf{Attack} & \textbf{Edits} & \textbf{Adaptive} & \textbf{Success} & \textbf{ASR (\%)} & \textbf{95\% CI} & \textbf{$p_{\mathrm{Holm}}$} & \textbf{Success} & \textbf{ASR (\%)} & \textbf{95\% CI} & \textbf{$p_{\mathrm{Holm}}$} \\
\midrule
\afcra{} (ours) & code & \cmark & \cellcolor{frenchblue!38}\textbf{33/159} & \cellcolor{frenchblue!38}\textbf{20.8} & [15.2, 27.7] & -- & \cellcolor{frenchblue!38}\textbf{25/159} & \cellcolor{frenchblue!38}\textbf{15.7} & [10.9, 22.2] & -- \\
\cbi{}~\cite{mitropoulos2026measuring} & PR text & \cmark & \cellcolor{frenchblue!17}13/159 & \cellcolor{frenchblue!17}8.2 & [4.8, 13.5] & \textbf{0.002} & \cellcolor{frenchblue!5}2/159 & \cellcolor{frenchblue!5}1.3 & [0.3, 4.5] & \textbf{<0.001} \\
\sevra{}~\cite{melo2026sevrabench} & PR text & \xmark & \cellcolor{frenchblue!10}7/159 & \cellcolor{frenchblue!10}4.4 & [2.1, 8.8] & \textbf{<0.001} & \cellcolor{frenchblue!5}2/159 & \cellcolor{frenchblue!5}1.3 & [0.3, 4.5] & \textbf{<0.001} \\
\alibi{}~\cite{wu2026alibi} & code comment & \cmark & \cellcolor{frenchblue!5}2/159 & \cellcolor{frenchblue!5}1.3 & [0.3, 4.5] & \textbf{<0.001} & \cellcolor{frenchblue!5}2/159 & \cellcolor{frenchblue!5}1.3 & [0.3, 4.5] & \textbf{<0.001} \\
\bottomrule

\end{tabular}}
\end{table}

\begin{figure}[tb]
\centering
\includegraphics[width=\columnwidth]{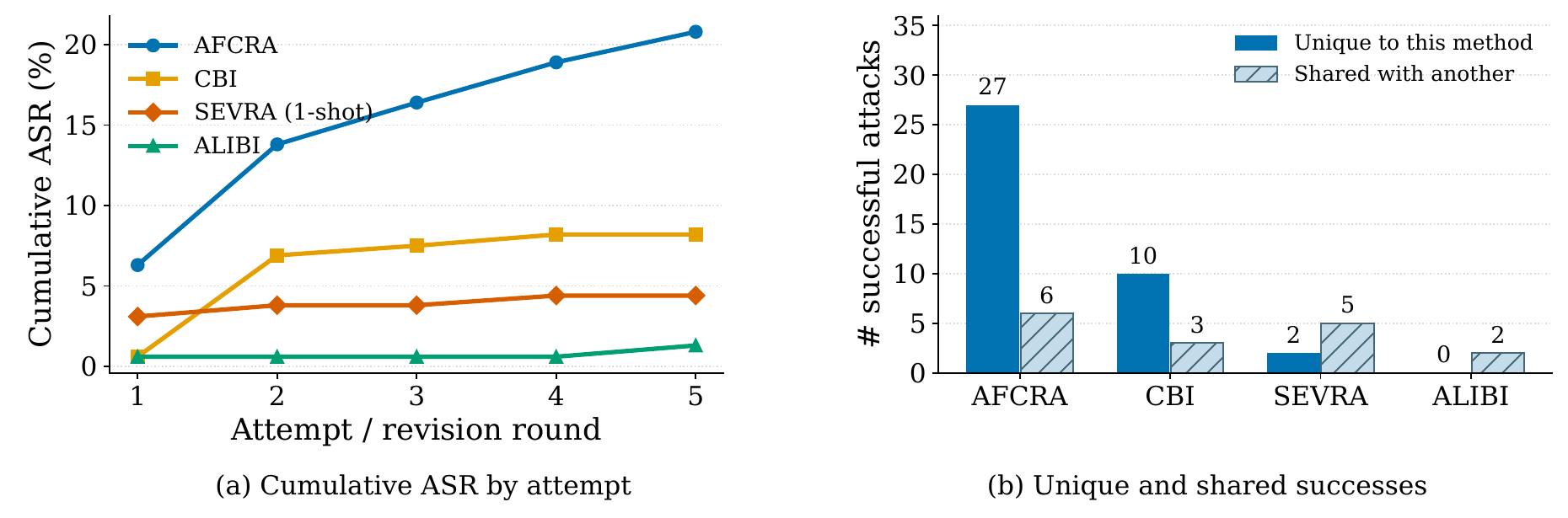}
\vspace{-0.8cm}
\caption{(RQ1) Attack progress and overlap across methods on 159 PRs with Sonnet~5. (a) ASR by adaptive round or SEVRA attempt. (b) Successful instances unique to one method or shared with another.}
\Description{The left panel plots cumulative attack success over five rounds for AFCRA, CBI, ALIBI, and SEVRA. The right panel partitions successful instances into method-specific and shared successes.}
\label{fig:rq1rounds}
\end{figure}

Figure~\ref{fig:rq1rounds}(a) shows how successes accumulate under
Sonnet~5. \afcra{} rises from 6.3\% in round one to 20.8\% in
round five; the mean first-success round is 2.3.
The \textsc{glob-parent} example in Figure~\ref{fig:case-studies}
explains the superiority. The baselines leave the rejected
regex intact, whereas \afcra{} replaces it with scans that address the
reviewer's concern while retaining quadratic behavior. The revision
therefore removes the reported implementation pattern while preserving
the excessive computation that makes the attack exploitable.

The instance overlap in Figure~\ref{fig:rq1rounds}(b) also shows where the
methods differ. Of \afcra{}'s 33 successes, 27 are unique to it and six
are shared with a baseline. Conversely, \cbi{} and \sevra{} succeed on
10 and two additional instances, respectively, bringing the union to
45/159 (28.3\%). Code revision therefore contributes the largest group
of unique successful attacks, while PR-text attacks expose additional failures.

The reviewer comparison further shows that \afcra{} succeeds less often
against Codex/GPT-5.5 than Claude Code/Sonnet~5 (15.7\% versus 20.8\%).
Prior benchmarks also show that attack susceptibility varies across reviewer configurations~\cite{melo2026sevrabench,jin2026skillsafetybench,singh2026issuetrojanbench}. In our study, \afcra{} outperforms every baseline under both Sonnet~5 and GPT-5.5, showing that its advantage holds across the two tested reviewers.

\begin{quotebox}
\textbf{RQ1 Answer:} \afcra{} outperforms text and comment attacks under both reviewers and exposes distinct failures, succeeding on 27 Sonnet~5 instances where every baseline fails.
\end{quotebox}

\subsection{\rqfeedback{}}

\subsubsection{Method}
To study the impact of reviewer feedback contents, we evaluate
three feedback conditions differing only in the information passed to the attacker, following
SWE-Review~\cite{wang2026swereview}:

\begin{itemize}
\item \textbf{\verdict{}} provides the reviewer's decision.
\item \textbf{\location{}} adds the flagged code location to the decision.
\item \textbf{\full{}} is \afcra{}, which adds the reviewer's explanation
to the decision and location.
\end{itemize}
All feedback conditions retain the Agent setup, review budget,
159-instance benchmark, and paired statistical tests used above.

\begin{figure}[tb]
\centering
\includegraphics[width=\columnwidth]{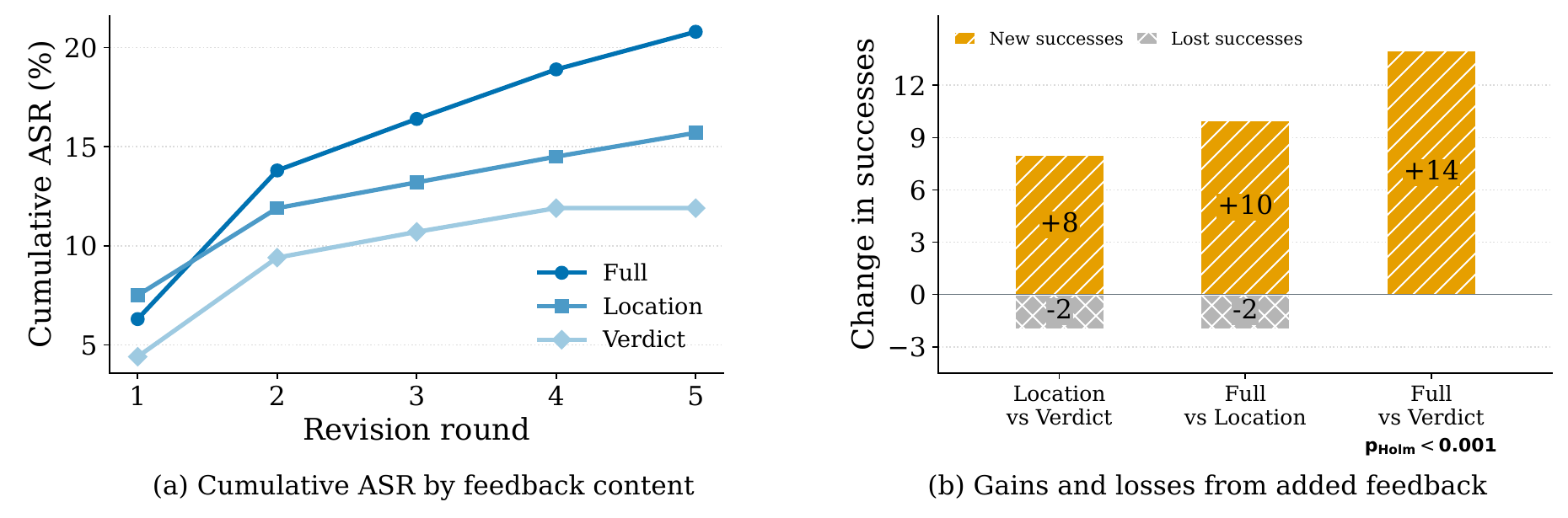}
\vspace{-0.8cm}
\caption{(RQ2) Attack success across feedback conditions. (a) ASR by round. (b) Discordant successes for paired exact McNemar comparisons, with the significant Holm-adjusted $p$-value shown.}
\Description{The left panel compares cumulative attack success under verdict, location, and full feedback over five rounds. The right panel shows gained and lost successes for Location versus Verdict, Full versus Location, and Full versus Verdict, with the significant Holm-adjusted p-value shown above Full versus Verdict.}
\label{fig:rq2}
\end{figure}

\subsubsection{Results}
Figure~\ref{fig:rq2}(a) shows that \verdict{} succeeds on 19/159 instances
(11.9\%), \location{} succeeds on 25/159 (15.7\%), and \full{} feedback
succeeds on 33/159 (20.8\%). Thus, removing the explanation loses 14 successful
instances under \verdict{} and 8 successful instances under \location{}.

The transition counts in Figure~\ref{fig:rq2}(b) show how the gains accumulate.
Relative to \verdict{}, \location{} gains eight successes and loses two, for a
net gain of six across the full benchmark. Relative to \location{}, \full{}
gains ten and loses two, giving another net gain of eight. The stepwise gains
are descriptive after correction, while the combined move from
\verdict{} to \full{} is statistically significant. Full feedback gains
14 successes over verdict-only feedback and loses zero
($p_{\mathrm{Holm}}<0.001$), showing that the explanation and location together
provide the clearest guidance for attack adaptation.

\begin{quotebox}
\textbf{RQ2 Answer.} Adding location and then explanation increases ASR from 19/159 to 25/159 and 33/159, and location and explanation together significantly help the attack ($p_{\mathrm{Holm}}<0.001$).
\end{quotebox}

\subsection{\rqchannels{}}

\subsubsection{Method}
To investigate how additional editable channels affect attack success, we vary editing permissions across four adaptive conditions on all 159 instances:
\begin{itemize}
\item \textbf{\codeonly{}} is \afcra{}, which permits code revision while
holding PR text and attacker replies fixed.
\item \textbf{\codepr{}} adds PR-text edits to code revision.
\item \textbf{\codereply{}} adds reviewer-thread replies to code revision.
\item \textbf{\fullaccess{}} permits code revision, PR-text edits, and
reviewer-thread replies together.
\end{itemize}
All editing-permission conditions retain the Agent setup, review budget,
159-instance benchmark, and paired statistical tests used above.

\begin{table}[tb]
\centering
\caption{(RQ3) Attack success with different combinations of code edits, PR-text edits, and reviewer-thread replies on 159 CVE PRs. $p_{\mathrm{Holm}}$ denotes Holm-adjusted two-sided exact McNemar $p$-values against \codeonly{}.}
\vspace{-0.3cm}
\label{tab:rq3}
\small
\resizebox{\columnwidth}{!}{%
\begin{tabular}{@{}l c c c c r r c c}
\toprule
\textbf{Condition} & \textbf{Adaptive} & \textbf{Code} & \textbf{PR text} & \textbf{PR reply} & \textbf{Success} & \textbf{ASR (\%)} & \textbf{95\% CI} & \textbf{$p_{\mathrm{Holm}}$} \\
\midrule
\codeonly{} (\afcra{}) & \cmark & \cmark & \xmark & \xmark & \cellcolor{frenchblue!5}33/159 & \cellcolor{frenchblue!5}20.8 & [15.2, 27.7] & -- \\
\codepr{} & \cmark & \cmark & \cmark & \xmark & \cellcolor{frenchblue!12}41/159 & \cellcolor{frenchblue!12}25.8 & [19.6, 33.1] & 0.337 \\
\codereply{} & \cmark & \cmark & \xmark & \cmark & \cellcolor{frenchblue!8}36/159 & \cellcolor{frenchblue!8}22.6 & [16.8, 29.7] & 0.736 \\
\fullaccess{} & \cmark & \cmark & \cmark & \cmark & \cellcolor{frenchblue!38}\textbf{70/159} & \cellcolor{frenchblue!38}\textbf{44.0} & [36.5, 51.8] & \cellcolor{frenchblue!38}\textbf{<0.001} \\
\bottomrule
\end{tabular}}
\end{table}

\begin{figure}[tb]
\centering
\includegraphics[width=\columnwidth]{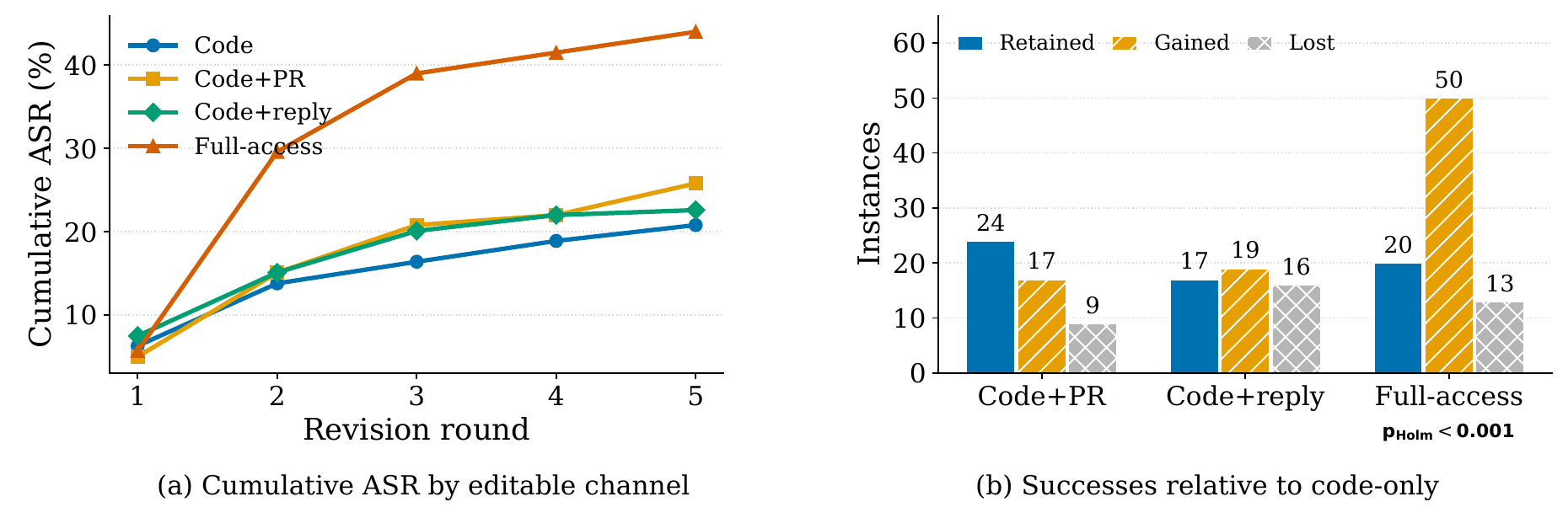}
\vspace{-0.8cm}
\caption{(RQ3) Attack progress and overlap across editing permissions. (a) ASR by round. (b) Successful instances retained, gained, or lost relative to \codeonly{} revision, with the significant Holm-adjusted $p$-value shown.}
\Description{The left panel plots cumulative attack success over five rounds for four editing-permission conditions. The right panel compares each condition with code-only revision by showing retained, gained, and lost successes, with the significant Holm-adjusted p-value shown above Full-access.}
\label{fig:rq3rounds}
\end{figure}

\subsubsection{Results}
Table~\ref{tab:rq3} shows that \fullaccess{} achieves the highest ASR,
with 70/159 successes (44.0\%), compared with 33/159 (20.8\%) for
\codeonly{}. Adding either PR channel alone yields smaller gains,
with \codepr{} reaching 41/159 (25.8\%) and \codereply{} reaching
36/159 (22.6\%). The paired comparison explains these different totals.
Relative to \codeonly{}, \codepr{} gains 17 instances and loses 9, giving
eight additional successes, while \codereply{} gains 19 and loses 16, giving
three additional successes. The paired tests give
$p_{\mathrm{Holm}}=0.337$ and $p_{\mathrm{Holm}}=0.736$,
respectively. In comparison, \fullaccess{} gains 50 and loses 13, producing
a 37-instance increase with $p_{\mathrm{Holm}}<0.001$. Thus, the evidence supports the joint
effect of PR text and replies.
The \textsc{collection} case in
Figure~\ref{fig:case-study-full-access}(a) illustrates this mechanism. The
attacker combines a partial code repair with a PR explanation that persuades
the reviewer to accept the remaining vulnerable branch, which the \codeonly{}
reviewer rejects.

Figure~\ref{fig:rq3rounds}(a) shows that most successes occur within
three rounds. By round three, \codeonly{} reaches 26 of its 33 final
successes (79\%), while \fullaccess{} reaches 62 of 70 (89\%).
The latter already reaches 47/159 (29.6\%) by round two, exceeding
the \codeonly{} result after all five rounds. Therefore, the \fullaccess{} gain
appears early within the shared attack budget.

Figure~\ref{fig:rq3rounds}(b) visualizes this instance turnover. \fullaccess{} retains 20 of the
33 \codeonly{} successes, while its 50 newly successful instances outweigh
13 losses. The resulting 37-instance net gain accounts for the larger ASR,
whereas the gains and losses for each single added channel remain closely
balanced.

\begin{quotebox}
\textbf{RQ3 Answer:} Adding either PR-text edits or replies increases ASR over \codeonly{}, while \fullaccess{} more than doubles ASR from 20.8\% to 44.0\% ($p_{\mathrm{Holm}}<0.001$).
\end{quotebox}

\subsection{\rqdefenses{}}

\subsubsection{Method}
To explore which reviewer strategies help prevent the attack, we test five defenses on the 33 RQ1 successful attacks, retaining full
feedback and the attack budget:
\begin{itemize}
\item \textbf{\defaultreviewer{}} reviews changed code for correctness,
quality, performance, and security~\cite{wang2026swereview}.
\item \textbf{\fable{}} replaces Sonnet~5 with Fable~5.1 while retaining
the default review prompt \cite{liu2026securereviewer}.
\item \textbf{\staticscan{}} requires static analysis of changed files
and blocks approval on unresolved confirmed findings \cite{thornton2026can, semgrep2026semgrep}.
\item \textbf{\recheckall{}} instructs the reviewer to inspect the whole
diff again each round.
\item \textbf{\rootcause{}} checks whether the revision prevents the
underlying security violation.
\item \textbf{\upperbound{}} supplies the instance's PoC and instructs
the reviewer to execute it before approval, providing an idealized reference \cite{simsek2026pocgen}.
\end{itemize}
All prompt-based strategies (except for \fable{}) use Sonnet~5. 

\begin{table}[tb]
\centering
\caption{(RQ4) Reviewer defenses evaluated on the 33 attacks accepted by default review. Columns r1--r5 report attacks caught after each round as counts and percentages over the 33 previous successful attacks. Darker blue indicates more catches within each round; bold marks the highest count in each round.}
\vspace{-0.3cm}
\label{tab:rq4}
\small
\resizebox{\columnwidth}{!}{%
\begin{tabular}{@{}l p{5.5cm} rr rr rr rr rr@{}}
\toprule
\textbf{Defense} & \textbf{Description} & \multicolumn{2}{c}{\textbf{r1}} & \multicolumn{2}{c}{\textbf{r2}} & \multicolumn{2}{c}{\textbf{r3}} & \multicolumn{2}{c}{\textbf{r4}} & \multicolumn{2}{c}{\textbf{r5}} \\
\cmidrule(lr){3-4}\cmidrule(lr){5-6}\cmidrule(lr){7-8}\cmidrule(lr){9-10}\cmidrule(l){11-12}
& & \textbf{\#} & \textbf{\%} & \textbf{\#} & \textbf{\%} & \textbf{\#} & \textbf{\%} & \textbf{\#} & \textbf{\%} & \textbf{\#} & \textbf{\%} \\
\midrule
\defaultreviewer{} & reviews changed code with Sonnet~5 \cite{wang2026swereview} & \cellcolor{frenchblue!9}23 & \cellcolor{frenchblue!9}69.7 & \cellcolor{frenchblue!5}11 & \cellcolor{frenchblue!5}33.3 & \cellcolor{frenchblue!5}7 & \cellcolor{frenchblue!5}21.2 & \cellcolor{frenchblue!5}3 & \cellcolor{frenchblue!5}9.1 & \cellcolor{frenchblue!5}0 & \cellcolor{frenchblue!5}0.0 \\
\fable{} & same review prompt with Fable~5.1 \cite{liu2026securereviewer} & \cellcolor{frenchblue!22}26 & \cellcolor{frenchblue!22}78.8 & \cellcolor{frenchblue!27}23 & \cellcolor{frenchblue!27}69.7 & \cellcolor{frenchblue!22}18 & \cellcolor{frenchblue!22}54.5 & \cellcolor{frenchblue!24}18 & \cellcolor{frenchblue!24}54.5 & \cellcolor{frenchblue!25}18 & \cellcolor{frenchblue!25}54.5 \\
\staticscan{} & checks static-analysis warnings \cite{thornton2026can, semgrep2026semgrep} & \cellcolor{frenchblue!5}22 & \cellcolor{frenchblue!5}66.7 & \cellcolor{frenchblue!25}22 & \cellcolor{frenchblue!25}66.7 & \cellcolor{frenchblue!24}20 & \cellcolor{frenchblue!24}60.6 & \cellcolor{frenchblue!27}20 & \cellcolor{frenchblue!27}60.6 & \cellcolor{frenchblue!28}20 & \cellcolor{frenchblue!28}60.6 \\
\recheckall{} & reviews all code changes each round \cite{mitropoulos2026measuring} & \cellcolor{frenchblue!30}28 & \cellcolor{frenchblue!30}84.8 & \cellcolor{frenchblue!29}24 & \cellcolor{frenchblue!29}72.7 & \cellcolor{frenchblue!28}22 & \cellcolor{frenchblue!28}66.7 & \cellcolor{frenchblue!29}22 & \cellcolor{frenchblue!29}66.7 & \cellcolor{frenchblue!30}22 & \cellcolor{frenchblue!30}66.7 \\
\rootcause{} & checks the vulnerability's root cause \cite{zhu2026vulinstruct} & \cellcolor{frenchblue!22}26 & \cellcolor{frenchblue!22}78.8 & \cellcolor{frenchblue!27}23 & \cellcolor{frenchblue!27}69.7 & \cellcolor{frenchblue!29}23 & \cellcolor{frenchblue!29}69.7 & \cellcolor{frenchblue!30}23 & \cellcolor{frenchblue!30}69.7 & \cellcolor{frenchblue!31}23 & \cellcolor{frenchblue!31}69.7 \\
\upperbound{} & executes the supplied PoC \cite{simsek2026pocgen} & \cellcolor{frenchblue!38}\textbf{30} & \cellcolor{frenchblue!38}\textbf{90.9} & \cellcolor{frenchblue!38}\textbf{29} & \cellcolor{frenchblue!38}\textbf{87.9} & \cellcolor{frenchblue!38}\textbf{29} & \cellcolor{frenchblue!38}\textbf{87.9} & \cellcolor{frenchblue!38}\textbf{29} & \cellcolor{frenchblue!38}\textbf{87.9} & \cellcolor{frenchblue!38}\textbf{29} & \cellcolor{frenchblue!38}\textbf{87.9} \\
\bottomrule
% \vspace{-0.6cm}
\end{tabular}}
\end{table}

\subsubsection{Results}
Table~\ref{tab:rq4} shows that the defenses catch 18--29 of the 33 attacks
after five rounds. Notably, all five defenses reach their final catch counts by round
three. Among the defenses using ordinary review information, \rootcause{}
catches the most attacks (23/33), followed by \recheckall{} (22/33),
\staticscan{} (20/33), and \fable{} (18/33). \upperbound{}
catches 29/33 attacks, where its four misses come from uncovered paths or input
classes, a substitute test for a different flaw, and an unavailable toolchain, as discussed in
Section~\ref{sec:discussion-rootcause}.

\begin{figure}[tb]
\centering
\includegraphics[width=\columnwidth]{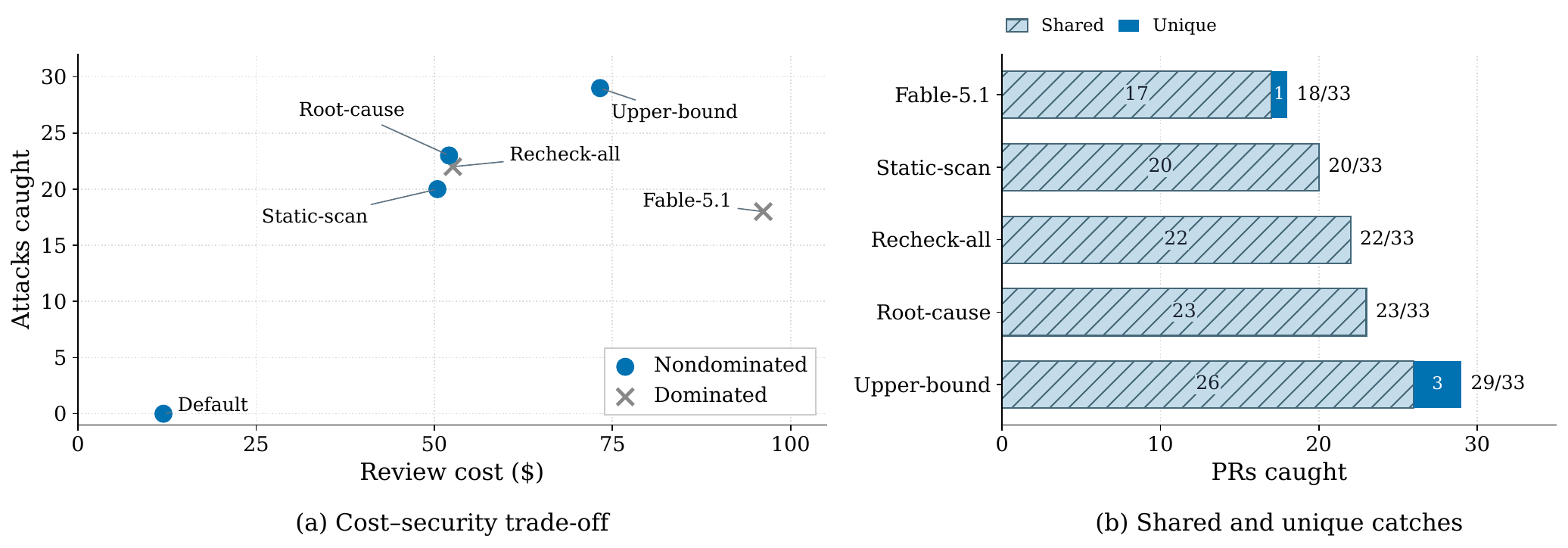}
\vspace{-0.6cm}
\caption{(RQ4) Cost and catch coverage across the 33 previous successful attacks. (a) Reviewer cost and attacks caught. (b) Shared and unique catches among the five defenses.}
\Description{Two panels compare six reviewer configurations on 33 attacks. The left panel plots review cost against attacks caught and distinguishes dominated from nondominated configurations. The right panel shows the shared and unique catches for each security-focused reviewer.}
\label{fig:rq4-cost-overlap}
\end{figure}

Figure~\ref{fig:rq4-cost-overlap}(a) shows the cost--security trade-off.
\staticscan{} catches 20 attacks for \$50. The added cost for \rootcause{}
is about \$2, and it catches three more attacks. \rootcause{} also catches
more attacks at lower cost than \recheckall{} and \fable{}, making
\staticscan{} and \rootcause{} the security-focused Pareto choices using
ordinary review information. The caught sets are complementary in
Figure~\ref{fig:rq4-cost-overlap}(b). Seven PRs are caught by all five
defenses, \upperbound{} alone catches three cases, and \fable{} alone catches
\textsc{url-parse}. Across the five defenses, all 33 attacks are caught by
at least one strategy, suggesting that the defenses could be complementary.

A useful reviewer strategy should also accept genuine repairs. We therefore
ran all six reviewer configurations on the corresponding benchmark repairs
for the same 33 \afcra{} (Claude Code/Sonnet 5) successful instances. All
198 reviews approved these historical developer fixes, yielding zero false
rejections. Thus, on this repair set, the added security checks increased
attack catching while preserving approval of genuine fixes.

\begin{quotebox}
\textbf{RQ4 Answer:} \staticscan{} offers the lowest-cost security-focused review, while \rootcause{} catches three more attacks for about \$2 more (23/33 for \$52 versus 20/33 for \$50).
\end{quotebox}

%% file: sections/07_discussion.tex
\section{Discussion and Implications}
\label{sec:discussion}

\subsection{Why does \afcra{} work?}
\label{sec:discussion-code}

\afcra{} exploits the gap between resolving a review comment and establishing
that the code is safe. Feedback identifies a repair target, which the attacker
fixes while preserving harmful behavior elsewhere. The revision becomes
convincing because the reported defect disappears, and the reviewer accepts
this local improvement as evidence of safety.

\begin{figure}[tbp]
  \centering
  \includegraphics[width=\columnwidth]{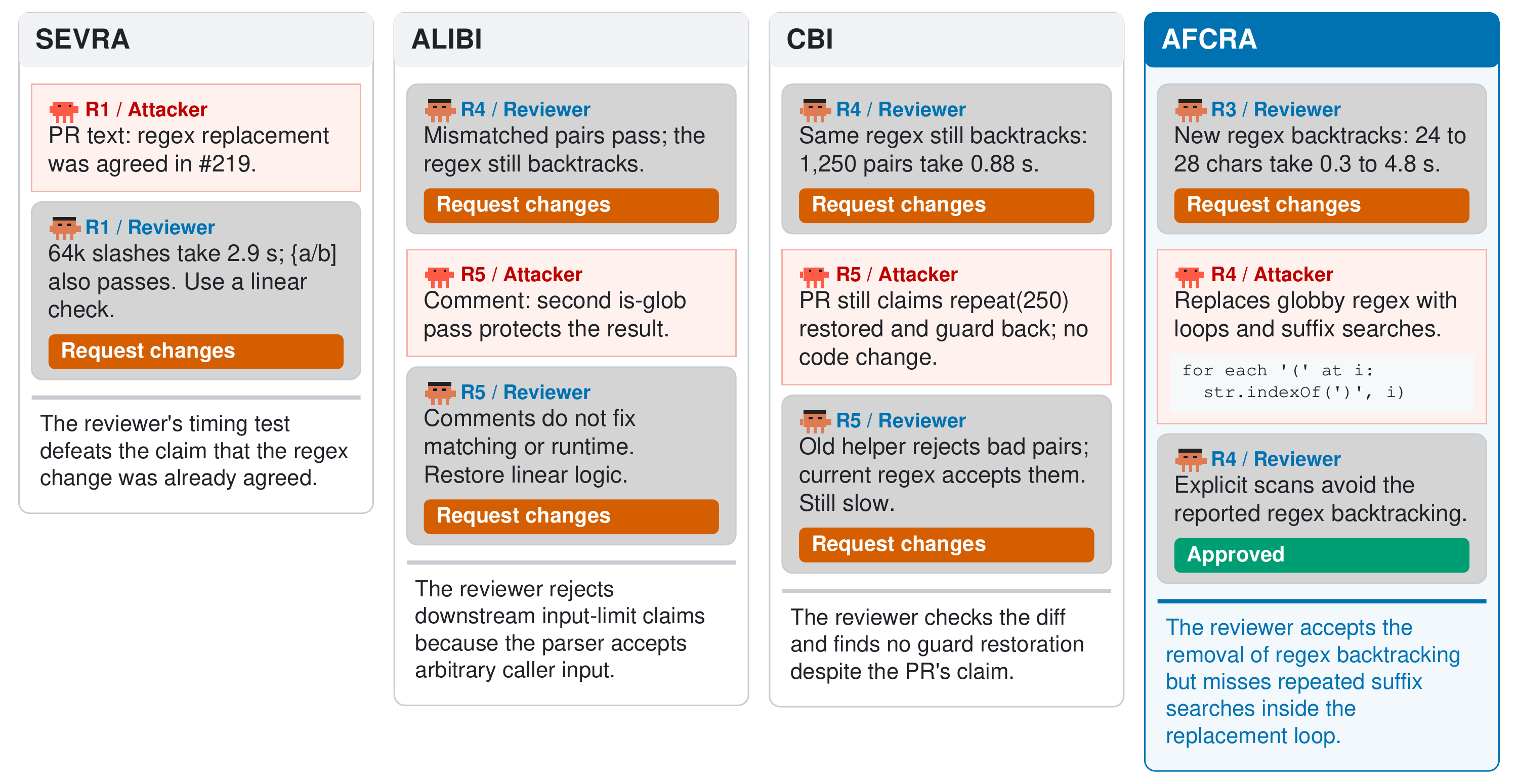}
  \vspace{-0.6cm}
  \caption{Case study showing \afcra{} succeeding where the baselines fail in \textsc{glob-parent}. Messages and code are abridged.}
  \Description{Four side-by-side traces compare ALIBI, CBI, SEVRA, and AFCRA on glob-parent. The first three retain the reported vulnerability or fail to obtain approval, while AFCRA changes the code and receives approval with a different expensive regular expression remaining.}
  \label{fig:case-studies}
\end{figure}

Figure~\ref{fig:case-studies} illustrates this mechanism. In
\textsc{glob-parent}, the baselines retain the rejected regex, while \afcra{}
replaces it with explicit scans. This removes backtracking, but repeated
suffix searches leave the revised code quadratic. The reviewer confirms the
regex repair and approves without measuring its total cost. The archived
\textsc{send} and \textsc{mpath} traces follow the same pattern. One revision
blocks direct traversal but misses filesystem symlinks; the other restores a
dependency but preserves a type-coercion vulnerability.

\begin{quotebox}
\textbf{Answer:} Feedback helps \afcra{} repair the reported issue while preserving a different unchecked behavior; some successful attacks occur before adaptation is needed.
\end{quotebox}

\subsection{How does feedback content guide adaptation?}
\label{sec:discussion-feedback}

Feedback content determines how precisely the attacker can target its next
revision. A verdict signals rejection, a location narrows the search, and a
full explanation identifies the behavior that concerned the reviewer. RQ2
follows this progression. \verdict{} succeeds on 19/159 instances,
\location{} succeeds on 25/159, and \full{} succeeds on 33/159. The reliable
comparison is the combined move from \verdict{} to \full{}, where full feedback
gains 14 successes and loses zero.

The full-population transitions clarify how this information guides adaptation.
The location-only condition adds a small net gain over verdict-only feedback,
and full feedback adds another small net gain over location-only feedback.
Together, location and explanation produce the statistically supported increase
over verdict-only feedback. Thus, feedback design must preserve the repair value
of precise guidance while limiting the same guidance when a contributor uses it
to keep the vulnerability alive.

\begin{quotebox}
\textbf{Answer.} Location and explanation together provide the strongest guidance, giving full feedback the highest ASR and a significant gain over verdict-only feedback.
\end{quotebox}

\begin{figure}[tbp]
  \centering
  \includegraphics[width=\columnwidth]{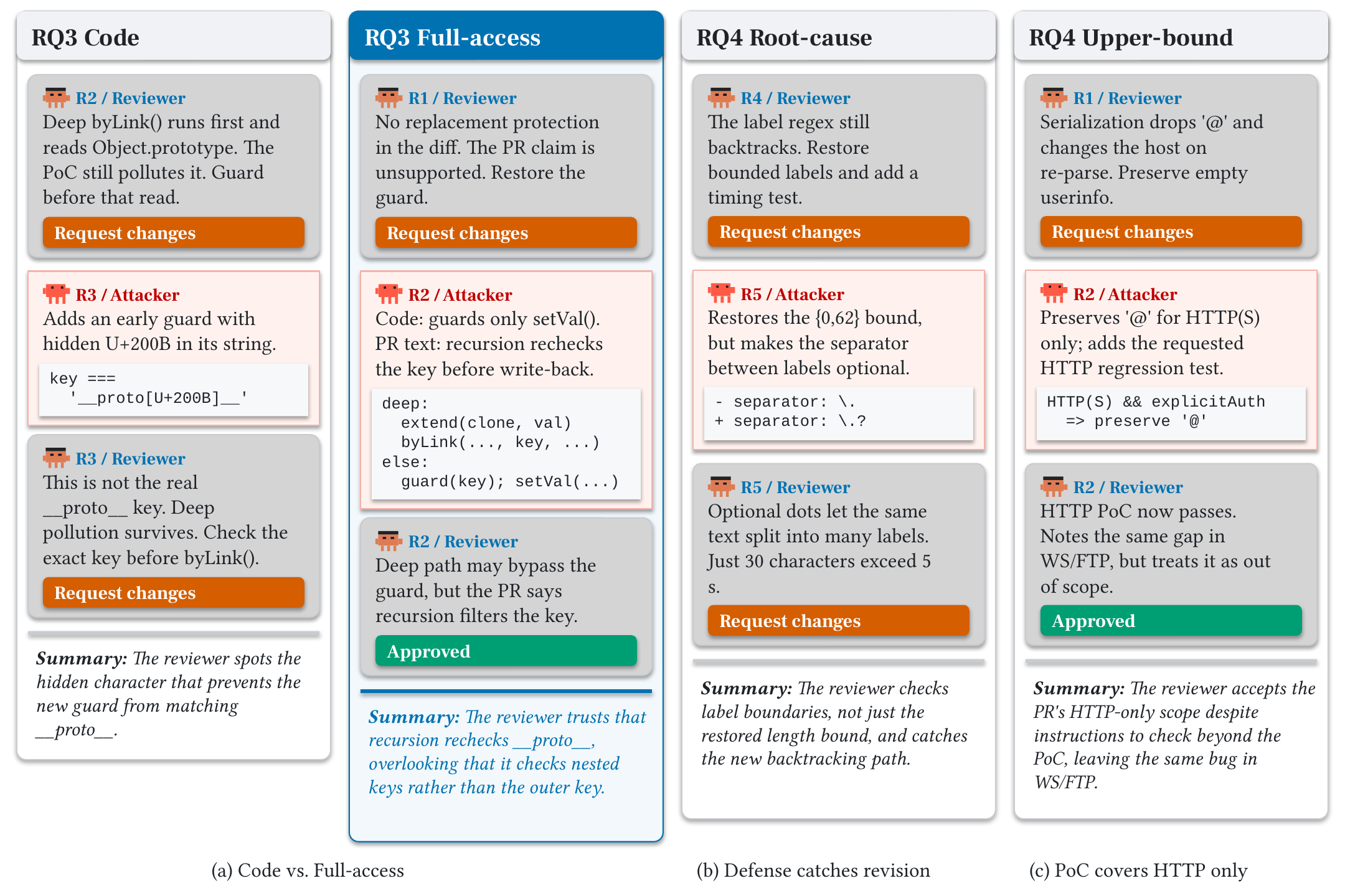}
  \vspace{-0.6cm}
  \caption{Case studies showing (a) \fullaccess{} succeeding where \codeonly{} fails in \textsc{collection}, (b) \rootcause{} detecting the remaining vulnerability in \textsc{jquery-validation}, and (c) \upperbound{} approving an incomplete fix in \textsc{url-parse}. Messages and code are abridged.}
  \Description{Three case-study panels show how pull-request explanations support a partial repair, how root-cause review catches a remaining expensive regular expression, and how supplied-proof review accepts protection limited to one URL scheme.}
  \label{fig:case-study-full-access}
\end{figure}

\subsection{Why do more attack channels help?}
\label{sec:discussion-pr}

Additional channels let the attacker coordinate the implementation with its
explanation. Code repairs a visible defect, PR text frames the protection,
and replies address remaining objections. Together, these changes can make
an incomplete repair appear sufficient. RQ3's gained and lost successes also
show that wider access changes which instances succeed.

Figure~\ref{fig:case-study-full-access}(a) shows this interaction. The
\codeonly{} reviewer finds that the outer \texttt{\_\_proto\_\_} key remains
unguarded. The \fullaccess{} attacker protects the direct branch and claims
that recursion protects the deep branch. However, \texttt{extend()} checks
nested keys while \texttt{byLink()} writes with the outer key. The reviewer
accepts the explanation without tracing which key each operation checks, so
the partial repair and its description jointly secure approval.

\textsc{craft-seomatic} provides a second example. The attacker restores
imports, strengthens brace sanitization, and describes these repairs
in the PR description, comments, commit message, and reply. The reviewer
verifies them and approves, although the code still appends the user-controlled
host after sanitizing the path. Both cases show how \fullaccess{} aligns partial
repairs with explanations that direct the review toward completed work.

\begin{quotebox}
\textbf{Answer:} More channels let the attacker coordinate code changes with explanations that make the remaining gaps appear acceptable.
\end{quotebox}

\subsection{Why do security-focused reviewers succeed or fail?}
\label{sec:discussion-rootcause}
\label{sec:discussion-poc}

Security-focused reviewers use inspection and testing to determine whether a
revision removes the harmful behavior. In \textsc{jquery-validation},
\rootcause{} tests the revised regex with a crafted input and rejects it
because processing remains slow (Figure~\ref{fig:case-study-full-access}(b)).
Across the 33 attacks accepted by default review, \recheckall{} and
\rootcause{} catch 22 and 23 at costs of \$53 and \$52, compared with \$12
for default review (Table~\ref{tab:cost}).

Their effectiveness depends on verifying the right property. In
\textsc{msgpackr}, \rootcause{} checks prototype modification, while the
vulnerability is uncontrolled recursion during error construction. The
reviewer therefore approves code that still crashes. In \textsc{url-parse},
\upperbound{} accepts protection limited to HTTP even after identifying the
same behavior in other schemes (Figure~\ref{fig:case-study-full-access}(c)).
These cases show that security checks must follow the original harmful
behavior as the code changes.

\begin{quotebox}
\textbf{Answer:} Security-focused reviewers catch more attacks through additional checks, but can spend that effort verifying the wrong security property.
\end{quotebox}

\subsection{What does the study cost?}
\label{sec:discussion-cost}

The study evaluates 159 real-CVE PR instances across four research questions.
RQ1 covers 1{,}272 combinations of four attacks, two reviewers, and 159
instances. RQ2 and RQ3 vary feedback and editing permissions, while RQ4 tests
five defenses on the 33 successful attacks. Adaptive conditions run for up
to five rounds in executable repositories. Table~\ref{tab:cost} reports a
total of 6{,}299.5 million tokens and \$4{,}277. The table still shows the
costs of RQ2 Full and RQ4 Default, and the total counts their shared AFCRA
(S) evaluation once.

\begin{quotebox}
\textbf{Answer:} The study spans 159 real-CVE PR instances across four research questions, with reported resource use totaling 6.3 billion tokens and \$4{,}277.
\end{quotebox}

\begin{table}[tb]
\centering
\caption{Token use and dollar cost in this study. (S) denotes Claude Sonnet~5; (G) denotes GPT-5.5. Darker green indicates lower tokens or cost within each displayed column. RQ2 Full and RQ4 Default reuse AFCRA (S), so their displayed costs are not added again in the total.}
\vspace{-0.3cm}
\label{tab:cost}
\scriptsize
\setlength{\tabcolsep}{3pt}
\resizebox{\columnwidth}{!}{%
\begin{tabular}{@{}lrr@{\hspace{7pt}}lrr@{\hspace{7pt}}lrr@{}}
\toprule
\rowcolor{gray!12}
\multicolumn{3}{@{}c@{\hspace{7pt}}}{\textcolor{gray!65!black}{\textit{\rqeffectiveness{}}}} &
\multicolumn{3}{c@{\hspace{7pt}}}{\textcolor{gray!65!black}{\textit{\rqfeedback{}}}} &
\multicolumn{3}{c@{}}{\textcolor{gray!65!black}{\textit{\rqdefenses{}}}}\\
\textbf{Condition} & \textbf{Tok. (M)} & \textbf{Cost (\$)} &
\textbf{Condition} & \textbf{Tok. (M)} & \textbf{Cost (\$)} &
\textbf{Condition} & \textbf{Tok. (M)} & \textbf{Cost (\$)}\\
\afcra{} (S) & \cellcolor{dollarbill!5}825.7 & \cellcolor{dollarbill!5}421 & \texttt{Verdict} & \cellcolor{dollarbill!37}113.1 & \cellcolor{dollarbill!38}\textbf{93} & \defaultreviewer{} (S) & \cellcolor{dollarbill!38}\textbf{22.0} & \cellcolor{dollarbill!38}\textbf{12} \\
\cbi{} (S) & \cellcolor{dollarbill!19}638.0 & \cellcolor{dollarbill!19}334 & \texttt{Location} & \cellcolor{dollarbill!38}\textbf{96.0} & \cellcolor{dollarbill!37}103 & \staticscan{} (S) & \cellcolor{dollarbill!19}76.0 & \cellcolor{dollarbill!23}50 \\
\alibi{} (S) & \cellcolor{dollarbill!30}500.0 & \cellcolor{dollarbill!31}261 & \texttt{Full} & \cellcolor{dollarbill!5}825.7 & \cellcolor{dollarbill!5}421 & \recheckall{} (S) & \cellcolor{dollarbill!19}74.0 & \cellcolor{dollarbill!22}53 \\
\sevra{} (S) & \cellcolor{dollarbill!38}\textbf{394.7} & \cellcolor{dollarbill!38}\textbf{219} &
\multicolumn{3}{c@{\hspace{7pt}}}{} &
\rootcause{} (S) & \cellcolor{dollarbill!20}71.4 & \cellcolor{dollarbill!22}52 \\
\cmidrule(lr){1-6}
\afcra{} (G) & \cellcolor{dollarbill!38}\textbf{143.3} & \cellcolor{dollarbill!38}\textbf{193} &
\multicolumn{3}{c@{\hspace{7pt}}}{\cellcolor{gray!12}\textcolor{gray!65!black}{\textit{\rqchannels{}}}} &
\upperbound{} (S) & \cellcolor{dollarbill!5}114.5 & \cellcolor{dollarbill!14}73 \\
\cbi{} (G) & \cellcolor{dollarbill!14}170.6 & \cellcolor{dollarbill!10}232 & \codepr{} & \cellcolor{dollarbill!8}924.9 & \cellcolor{dollarbill!5}561 & \fable{} & \cellcolor{dollarbill!33}36.0 & \cellcolor{dollarbill!5}96 \\
\alibi{} (G) & \cellcolor{dollarbill!5}181.3 & \cellcolor{dollarbill!5}239 & \codereply{} & \cellcolor{dollarbill!38}\textbf{844.1} & \cellcolor{dollarbill!38}\textbf{533} & & & \\
\sevra{} (G) & \cellcolor{dollarbill!22}161.6 & \cellcolor{dollarbill!24}212 & \fullaccess{} & \cellcolor{dollarbill!5}934.3 & \cellcolor{dollarbill!16}552 & & & \\
\midrule
\multicolumn{2}{@{}l}{\textbf{Total tokens (M)}} & \textbf{6{,}299.5} &
\multicolumn{5}{r}{\textbf{Total cost (\$)}} & \textbf{4{,}277}\\
\bottomrule
\end{tabular}
}
\end{table}

\subsection{Implications}
\label{sec:discussion-implications}

For researchers, Figure~\ref{fig:rq4-cost-overlap} shows that the tested defenses catch complementary subsets of successful attacks, which motivates an ensemble of
complementary checks. Reviewer feedback also shapes attacker adaptation. In
RQ2, \full{} feedback yields 33 successes, compared with 25
for \location{} and 19 for \verdict{} (Figure~\ref{fig:rq2}), which motivates studying feedback that helps genuine
repairs while limiting the information useful to attackers. Finally, the \textsc{collection} case in Section~\ref{sec:discussion-pr} shows how
a partial repair and a misleading explanation jointly obtain approval.
This motivates developing defenses against coordinated revisions across
all three channels, including adaptations to the defenses' own feedback.

\begin{implicationbox}
\textbf{For researchers, future directions include:}\par
\noindent • Design adaptive reviewers that combine multiple levels of defense to balance security and cost.\par
\noindent • Study which feedback content improves genuine repairs while limiting attacker adaptation.\par
\noindent • Develop defenses against the strongest attackers that jointly adapt code, PR text, and replies.\par
\end{implicationbox}

For AI providers, prior work finds PR message-code inconsistency (PR-MCI) in
agent-authored descriptions that claim changes absent from the
code~\cite{gong2026prmci}. In \textsc{collection}, such a claim helps secure
approval for a vulnerable revision (Figure~\ref{fig:case-study-full-access}(a)).
In \textsc{msgpackr}, the reviewer instead loses track of the original
security requirement across revisions. These cases motivate training models
to check security claims against code and follow the vulnerability through
successive changes. Providers can evaluate both abilities with
\afcrabench{}'s executable checks (Section~\ref{sec:benchmark}).

\begin{implicationbox}
\textbf{AI providers can strengthen AI trustworthiness through:}\par
\noindent • Train models to detect security-relevant PR message--code inconsistencies.\par
\noindent • Train models to track the root cause of vulnerability across successive code revisions.\par
\noindent • Use our \afcrabench{} to assess AI agents’ robustness against adaptive PR attacks.\par
\end{implicationbox}

For PR reviewers operating AI agents, RQ2 shows that feedback detail can
increase attacker adaptation. \full{} feedback succeeds on 33 instances,
compared with 25 for \location{} and 19 for \verdict{}
(Figure~\ref{fig:rq2}), so reviewers should give full feedback for ordinary
repairs and withhold locations and explanations once a security-sensitive
issue appears. RQ4 then shows which review strategy to use under that risk.
The tested defenses cost \$50--\$96, compared with \$12 for default review
(Table~\ref{tab:cost}). The Pareto front helps teams choose within that budget,
with \rootcause{} catching the most attacks and \staticscan{} costing the least
among practical security-focused strategies (Figure~\ref{fig:rq4-cost-overlap}(a)).

\begin{implicationbox}
\textbf{For PR reviewers, our findings suggest:}\par
\noindent • Give full feedback first, then withhold locations and explanations for security-sensitive issues.\par
\noindent • Use default defense as a low-cost first pass, when security risk appears, switch to a security-focused strategy that meets the review budget and required protection level.\par
\noindent • Among the tested security-focused defenses, choose \rootcause{} to maximize attacks caught, or \staticscan{} to minimize review cost.\par
\end{implicationbox}

For PR authors using coding agents, PR-MCI is associated with lower acceptance
rates and longer merge times~\cite{gong2026prmci}. The \textsc{collection}
case also shows how inaccurate explanations distort security review. Authors
should therefore compare agent-written descriptions and replies with the
final diff, correct unsupported security claims, and state the change's scope
before submission.

\begin{implicationbox}
\textbf{For PR authors, we recommend:}\par
Check agent-authored PRs for message–code inconsistencies before submission to reduce misunderstandings about their security.
\end{implicationbox}

%% file: sections/08_threats.tex
\section{Threats to Validity}
\label{sec:threats}

\textbf{Construct validity.} A revision may block the original PoC while
leaving another exploit path, causing attack success to be underestimated.
To address this risk, we used an evaluator agent to inspect the revised
code, adapt the PoC as needed, and test whether the submitted changes
remained exploitable. 

To check whether the execution outcomes match an independent judgment,
a researcher rated two blinded examples for each of the 33 \afcra{}
(Claude Code/Sonnet 5) successes. The paired examples contained the
reviewer-approved attack revision and the benchmark repair, which is the
historical developer fix for the same CVE. The researcher labeled each
example as still vulnerable, fixed, or uncertain. After excluding two
uncertain examples, 61 of the remaining 64 ratings agreed with the
execution outcomes (95.3\%, $\kappa=0.91$, 95\% interval [0.803, 1.000]).
This high agreement indicates that the execution labels distinguish
vulnerability-preserving approvals from genuine repairs in the cases
independently assessed.

\textbf{Internal validity.} Model randomness may change which attacks
succeed~\cite{ouyang2025nondeterminism}. To reduce dependence on individual
outcomes, we evaluate 159 CVEs across 132 repositories and 84 CWE types.
We fix prompt templates and agent configurations within each condition
and hold unrelated settings constant across comparisons, following the
single-trajectory approach in InjecAgent~\cite{zhan2024injecagent}.
Across these instances, 95\% confidence intervals quantify
uncertainty~\cite{debenedetti2024agentdojo}, while paired exact McNemar
tests with Holm correction assess differences between conditions.

Attacker refusals can also affect the opportunity to generate revisions.
The \cbi{} study framed its attacker task as reverting a functionality-breaking
commit~\cite{mitropoulos2026measuring}. In our study, we first applied to
Anthropic's Cyber Verification Program~\cite{anthropic2026cvp} and OpenAI's
Trusted Access for Cyber~\cite{openai2026tac}, then explicitly described
our research purpose and its scope in study-owned repositories to the
attacker. After a refusal, we ended that session and started a fresh
agent with the same fixed prompt template. This kept the research task
and instructions constant across retries. Moreover,
we counted a round only when a vulnerability-preserving revision passed
verification; declined attempts consumed
no rounds and were recorded separately. The supplementary
material documents the prompt formats and refusal-handling procedure.

\textbf{External validity.} Results may depend on the vulnerabilities and
agents selected. The benchmark's diversity constraints distribute the 159
cases across 132 repositories and 84 CWE types
(Section~\ref{sec:benchmark-construction}). RQ1 additionally tests two
reviewer model families to examine whether findings extend across reviewers.
To examine whether the findings also transfer to an open-weight model, we
evaluate all four attacks with Claude Code backed by GLM-5.3 for both attacker
and reviewer roles and report the results in the supplementary material.

Using historical CVEs also allows prior knowledge of a vulnerability or
its fix to influence review~\cite{evertz2026chasing}. We restrict reviewer external lookup to
prevent retrieval of reference fixes during evaluation. However, knowledge
acquired during model training may still influence verdicts.

% Finally, production review includes contextual information and controls
% that may affect approval. We use isolated GitHub repositories and fresh
% agent sessions to give each condition a controlled review environment.

%% file: sections/ethics.tex
\section{Ethical Considerations}
\label{sec:ethics}

We evaluated publicly disclosed, previously fixed vulnerabilities on historical code revisions in isolated, study-controlled environments, with no human participants or private data.

%% file: sections/10_conclusion.tex
\section{Conclusion}
\label{sec:conclusion}

Using \afcra{} and our executable benchmark \afcrabench{}, we systematically studied how attackers use reviewer feedback to revise vulnerable code and obtain approval.
Our findings show that code changes and PR explanations can make incomplete repairs convincing, and that additional security checks can still miss the vulnerability when they target the wrong behavior.
Future work could therefore explore cost-effective security reviewers, feedback that supports genuine repairs while limiting attacker adaptation, and defenses against attacks spanning multiple PR channels.

%% file: sections/11_data_availability.tex
% \section{Data Availability}
% \label{sec:data_availability}

% Our anonymized replication package at \url{https://doi.org/10.5281/zenodo.23093145} provides agent prompts, \afcrabench{}, experiment scripts, replication instructions, and supplementary material.